\documentstyle[12pt]{article}
\begin{document}
\title{Time-Varying Fine-Structure Constant and Quantum SuperStrings}
\author{B.G. Sidharth$^*$\\ Centre for Applicable Mathematics \& Computer Sciences\\
B.M. Birla Science Centre, Hyderabad 500 063 (India)}
\date{}
\maketitle
\footnotetext{E-mail:birlasc@hd1.vsnl.net.in}
\begin{abstract}
In this note we observe that Webb and co-workers' observation of an evolving
fine structure constant has since been reconfirmed. We also confirm Kuhne's
contention that this implies a cosmological constant and exhibit a cosmological
mechanism for Webb's result. We also see how this provides a mechanism to
reach elementary particle scales from the Quantum Superstring scale.
\end{abstract}
Kuhne\cite{r1} suggested sometime ago that the then
preliminary results of Web et al\cite{r2} that there was a slow cosmic
evolution of the fine struture constant $\alpha$, could be explained in
terms of a cosmology with a cosmological constant. Since then Web and
co-workers have reconfirmed their initial observations\cite{r3}, and the
acceleration of the universe, implying a cosmological constant has also
been repeatedly confirmed\cite{r4}. In this note we confirm Kuhne's conclusion
and provide a mechanism for the variation of $\alpha$.\\
Kuhne derives the following equation,
\begin{equation}
\frac{\dot \alpha_z}{\alpha_z} = \alpha_z \frac{\dot H_z}{H_z},\label{e1}
\end{equation}
from a well known equation of Teller, where $H$ is the Hubble constant, $z$ denotes
the value of the parameterr at the cosmological redshift $z$ and the dot
represents the derivative with respect to time. He then compares (\ref{e1})
with the finding of Web and co-workers, viz.,
\begin{equation}
\frac{\dot \alpha_z}{\alpha_z} \approx -1 \times 10^{-5} H_z.\label{e2}
\end{equation}
to exhibit the cosmological constant.\\
We consider the above results in the light of fluctuational cosmology\cite{r5,r6,r7}.
In this cosmology, $\sqrt{N}$ particles are fluctuationally created, somewhat
on the lines suggested by Hayakawa\cite{r8,r9}-and this happens within the
minimum unphysical time interval, the Compton time $\tau$. Such a cosmology is
not only consistent with observation, but it also deduces from the theory the
otherwise empirical Large Number Relations of Dirac as also the mysterious empirical
Weinberg relation between the pion mass and the Hubble Constant. The cosmology
also predicted an ever expanding and accelerating universe, as indeed was
discovered shortly thereafter and reconfirmed since (Cf. also \cite{r10}). In
this cosmological scheme, the cosmological constant $\Lambda$ is given by,
\begin{equation}
\Lambda \approx \dot H \leq - 0(H^2)\label{e3}
\end{equation}
In what follows, equalities are approximate equalities in the order of magnitude
sense.
We now observe that substitution of (\ref{e3}) in (\ref{e1}) gives
\begin{equation}
\frac{\dot \alpha_z}{\alpha_z} = \beta H_z\label{e4}
\end{equation}
where $\beta < - \alpha_z < - 10^{-2}$.\\
It can immediately be seen that (\ref{e4}) is compatible with (\ref{e2}).\\
We finally give the derivation of (\ref{e2}) in the above context wherein, as the
number of particles in the universe increases with time, we go from the Planck
scale to the Compton scale. It is known that this Compton length, due to
Zitterbewegung causes a correction to the electrostatic potential which an
orbiting electron experiences, rather like the Darwin term\cite{r11}.\\
Infact we have
$$\langle \delta V \rangle = \langle V (\vec r + \delta \vec r)\rangle - V
\langle (\vec r )\rangle$$
$$= \langle \delta r \frac{\partial V}{\partial r} + \frac{1}{2} \sum_{\imath j}
\delta r_\imath \delta r_j \frac{\partial^2 V}{\partial r_\imath \partial r_j} \rangle$$
\begin{equation}
\approx 0(1) \delta r^2 \nabla^2 V\label{e5}
\end{equation}
Remembering that $V = e^2/r$ where $r \sim 10^{-8}cm$,
from (\ref{e5}) it follows that if $\delta r \sim l$, the Compton wavelength
then
\begin{equation}
\frac{\Delta \alpha}{\alpha} \sim 10^{-5}\label{e6}
\end{equation}
where $\Delta \alpha$ is the change
in the fine structure constant from the early universe. (\ref{e6}) is an
equivalent form of (\ref{e2}) (Cf.ref.\cite{r1}), and is the result originally
obtained by Webb et al (Cf.refs.\cite{r2,r3}).\\
We now consider briefly Quantum Superstrings.\\
One of the criticisms levelled against Quantum Superstrings (QSS) Theory is
that it is not verifiable, as it deals with Planck scale phenomena. Indeed
't Hooft has gone so far as to say that it is not even a theory\cite{r12}.
The question is, are there any effects which can be observed at scales
corresponding to elementary particles? For example could we go from QSS
to the Compton scale? We will now discuss briefly, exactly such a mechanism.\\
As is well known, at the scale of QSS we have a non commutative geometry
\cite{r13,r14},
\begin{equation}
[x,y] \approx 0(l^2), [x,p_x] = \hbar [1+0(l^2)] etc.\label{e7}
\end{equation}
Another way of looking at (\ref{e7}) is that there is a generalised Uncertainity
Principle in operation at this scale (\cite{r15} - \cite{r20})
\begin{equation}
\Delta x \approx \frac{\hbar}{\Delta p} + l^2_p \frac{\Delta p}{\hbar}\label{e8}
\end{equation}
$l_p$ being the Planck length.\\
The first term on the right side of equation (\ref{e8}) gives the usual Uncertainity
relation, while the second term represents the duality effect - as we go down
to the Planck scale we infact are lead to the larger scale represented by the
second term (Cf.ref.\cite{r15,r21,r22}).\\
Before proceeding further, it may be remarked that the non commutativee geometry
in (\ref{e7}) is a manifestation of the non zero spatial extension of the
strings\cite{r23}.\\
We now come to a mechanism by which the Compton scale arises quite naturally
in the above context.
As was pointed out a long time ago by Hayakawa\cite{r24}, the fluctuation in
the mass of a typical elementary particle due to the fluctuation of the particle
number, which is $\sim \sqrt{N}, N \sim 10^{80}$ being the number of elementary particles in
the universe, is given by
$$\frac{G\sqrt{N}m^2}{c^2R}$$
Using now the Uncertainity Principle, we get,
\begin{equation}
(\Delta mc^2)T = \frac{G\sqrt{N}m^2}{R} T = \frac{G\sqrt{N}m^2}{c}\label{e9}
\end{equation}
$T$ being the age of the universe and $R$ its radius which equals $cT$.
It can be easily seen that the right side of (\ref{e9}) equals
$\hbar$! That is, we have,
\begin{equation}
\hbar \approx \frac{G\sqrt{N}m^2}{c}\label{e10}
\end{equation}
Equation (\ref{e10}) immediately gives us the Compton wavelength. Interestingly
a similar line of reasoning leads to a fluctuational model of cosmology, which
explains the many so called Large Number coincidences as also Weinberg's apparently
mysterious empirical relation between the pion mass and the Hubble
Constant, from theory, and moreover
predicts an ever expanding accelerating universe, as indeed has been repeatedly
verified in the past few years\cite{r25,r26}.\\
We now observe that $n \equiv \sqrt{N} \sim 10^{40}$. In the context of (\ref{e8}),
let us consider a secondary fluctuation $\sim \sqrt{n}$ of Planck scale particles.
Then the second term of (\ref{e8}) gives
\begin{equation}
\Delta x = l^2_p \cdot \frac{\sqrt{n}m_p c}{\hbar} = l_p \sqrt{n}\label{e11},
\end{equation}
$m_p$ being the Planck mass. The right side of (\ref{e11}) is precisely the Compton wavelength $l_\pi$ of a
typical elementary particle, viz., the pion. Immediately we recover therefrom
two other consistent relations for the mass of the pion and its Compton time
viz.,
$$m_\pi = \frac{m_p}{\sqrt{n}}, t_\pi = t_p \sqrt{n}$$
As mentioned after (\ref{e8}), this is the well known duality effect - as we
go down to the Planck scale, we end up at the Compton scale.\\
Thus the QSS scale and the Compton scale are related by the duality relation.\\
We finally make the following remark. In the light of the above considerations
leading to (\ref{e9}) and (\ref{e11}), it is interesting that we can consider the
energy of a Planck mass to be the gravitational energy of the fluctuational
$\sqrt{n}$ Planck masses within the Compton length $l_\pi$. That is, we should
have,
$$\frac{G\sqrt{n}m^2_p}{l_\pi} \approx m_p c^2.$$
This is indeed so.

\end{document}